\documentclass{kluwer}    

\usepackage{epsfig}   
\begin{document}
                 
\begin{article}
\begin{opening}         
\title{Are galaxies shy ?} 
\author{D. \surname{Elbaz}}
\runningauthor{D. Elbaz}
\runningtitle{Are galaxies shy ?}
\institute{CEA/DSM/DAPNIA, Service d'Astrophysique,
 F-91191 Gif-sur-Yvette C\'edex, France}

\begin{abstract}
\end{abstract}
\end{opening}           

\section{Introduction}  
Until 1996, there was little evidence that most galaxies were ``shy'',
i.e. that they would hide their stars behind a veil of dust and turn
red when forming stars, radiating the bulk of their luminosity in the
infrared (IR) at a given epoch of their history. Ten years before,
IRAS had unveiled a population of luminous IR galaxies exhibiting such
a ``shy'' behavior, the so-called LIGs and ULIGs (with 12$\geq
log_{10}\left(L_{\rm IR}/L_{\odot}\right)\geq$ 11 and
$log_{10}\left(L_{\rm IR}/L_{\odot}\right)\geq$ 12 respectively),
which are responsible for the shape of the bolometric luminosity
function of local galaxies above $\sim$ 10$^{11}~L_{\odot}$ (Sanders
\& Mirabel 1996). But integrated over the whole local luminosity
function, LIGs and ULIGs only produce $\sim$ 2\,$\%$ of the total
integrated luminosity and overall only $\sim$ 30\,$\%$ of the
bolometric luminosity of local galaxies is radiated in the IR above
$\lambda \sim$ 5\,$\mu$m. The discovery of an extragalactic background
in the IR at least as large as the UV-optical-near IR one, the
so-called cosmic infrared background (CIRB), with the COBE satellite
(Puget et al. 1996, see references in Elbaz et al. 2002b) implied that
shyness must have been more common among galaxies in the past than it
is today. This was confirmed with the detection of an excess of faint
mid IR (MIR) galaxies by ISOCAM onboard ISO (Elbaz et al. 1999), as
well as in the far IR (FIR) with ISOPHOT onboard ISO (Dole et
al. 2001) and in the sub-millimeter with SCUBA at the JCMT (see Smail
et al. 2001). This excess is relative to expectations based on
galaxies in the local universe. It implies that galaxies were more
luminous in the IR regime and/or more numerous in the past (Chary \&
Elbaz 2001, Franceschini et al. 2001).

\begin{figure} 
\centerline{\psfig{file=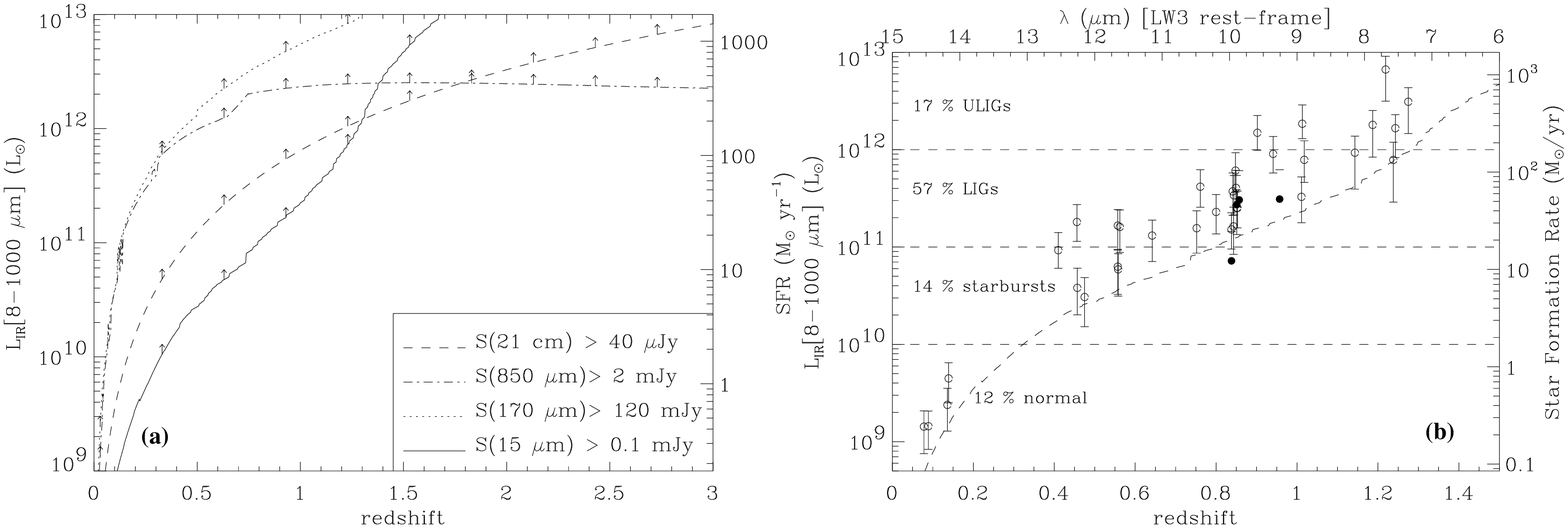,width=12cm}}
\caption[]{{\bf a)} IR luminosity (left axis) and SFR (right axis)
corresponding to the sensitivity limits of ISOCAM (15\,$\mu$m, plain
line), ISOPHOT (170\,$\mu$m, dotted line), SCUBA (850\,$\mu$m,
dot-dashed line) and VLA/WSRT (21 cm, dashed line, using the radio-FIR
correlation) as a function of redshift. K-correction from the library
of template SEDs from Chary \& Elbaz (2001). {\bf b)} $L_{\rm
IR}$[8-1000\,$\mu$m] and SFR versus redshift and $\lambda_{\rm
rest-frame}$ for the HDFN galaxies detected above a 15\,$\mu$m
completeness limit of 0.1 mJy (dashed line). Filled dots: 5 AGNs (left
axis only). }
\label{FIG:sfr}
\end{figure}

\section{Mid infrared as a star formation indicator}  

Chary \& Elbaz (2001) and Elbaz et al. (2002b) demonstrated that the
MIR luminosity of local galaxies is correlated with their integrated
IR luminosity (8-1000\,$\mu$m). Hence MIR flux densities can be
converted into $L_{\rm IR}$ and used to compute star formation rates
(SFR).  The sensitivities of the deepest surveys performed in the MIR
(0.1 mJy at 15\,$\mu$m), FIR (120 mJy at 170\,$\mu$m) and
sub-millimeter (2 mJy at 850\,$\mu$m) with ISOCAM, ISOPHOT and SCUBA
and in the radio (40\,$\mu$Jy at 1.4 GHz, i.e. 21 cm, with the VLA and
WSRT) to IR galaxies are compared in Fig.~\ref{FIG:sfr}a as a function
of redshift (see also Elbaz et al. 2002b). Fig.~\ref{FIG:sfr}a shows
that ISOCAM was the most sensitive instrument among the four selected
and that it was able to detect nearly all luminous IR galaxies below
$z\sim$ 1. A similar result is obtained using either the proto-typical
spectral energy distribution (SED) of M 82 or the library of 100
template SEDs from Chary \& Elbaz (2001) constructed to reproduce the
correlations between MIR-FIR and sub-millimeter luminosities of local
galaxies.

An indication that the SEDs in the IR of distant galaxies ressemble
local ones comes from the distant ``clone'' of Arp 220 serendipitously
discovered in the field of a QSO (PC 1643+4631). This galaxy, HR10
($z=$ 1.44) known as an extremely red object (ERO) was detected in the
radio, MIR and sub-millimeter with a SFR around 1000 M$_{\odot}$
yr$^{-1}$ (see Elbaz et al. 2002a and references therein).

The spatial resolution (4 arcsec PSF FWHM) of ISOCAM provided the
possibility to identify rather easily optical counterparts to these
galaxies and to determine their redshift. Due to limited telescope
time allocation, their redshift distribution was inferred from a
sub-sample of galaxies in the Hubble Deep Field North (HDFN, Aussel et
al. 1999) and their luminosities and star formation rates are
presented in the Fig.~\ref{FIG:sfr}b. About 75\,$\%$ of the galaxies
brighter than about 0.1 mJy at 15\,$\mu$m, and responsible for the
steep slope of the number counts, belong to the class of LIGs
($\sim$55\,$\%$) and ULIGs ($\sim$20\,$\%$). Their redshifts spread
over the $z=$ 0.5-1.3 range with a median around ${\bar z}=$ 0.7-0.8.

The fraction of IR light produced by active nuclei was computed from
the cross-correlation of ISOCAM with the deepest X-ray surveys from
the Chandra and XMM-Newton observatories in the HDFN (41 MIR galaxies)
and Lockman Hole (103 MIR galaxies) respectively. Less than 20\,$\%$
of the ISOCAM galaxies appear to be dominated by an AGN at 15\,$\mu$m
((12$\pm$5)\,$\%$ in the HDFN and (13$\pm$4)\,$\%$ in the Lockman
Hole, see Fadda et al. 2001).

\begin{figure} 
\centerline{\psfig{file=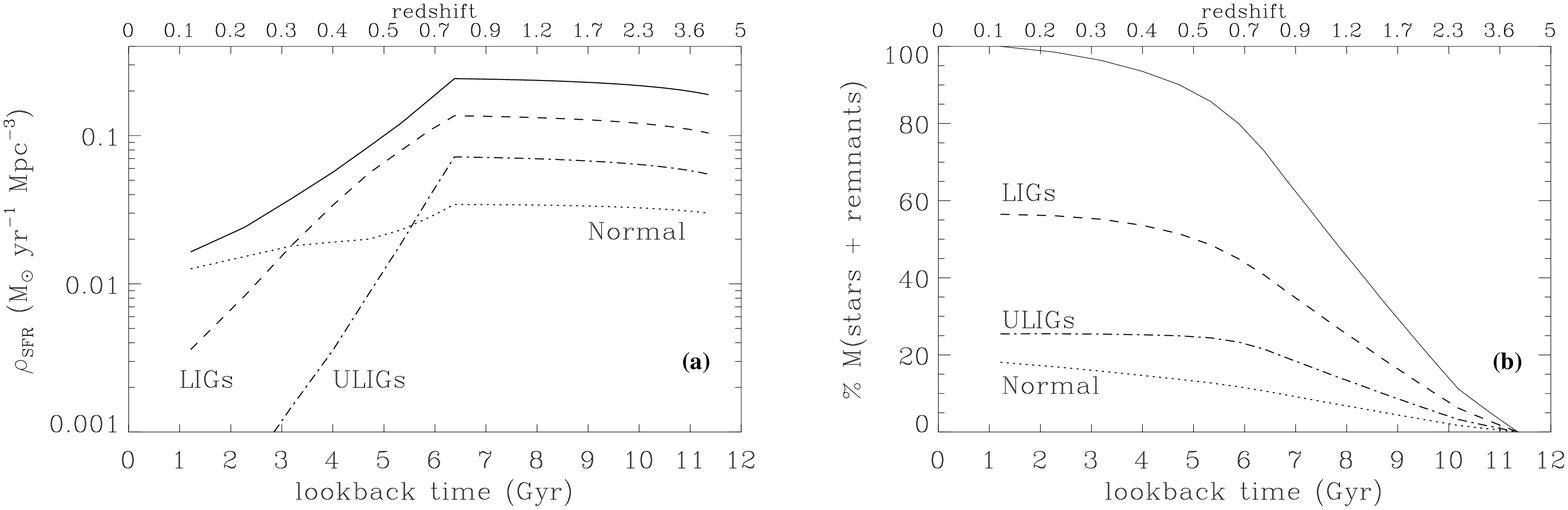,width=12cm}}
\caption[]{{\bf a)} Cosmic density of star formation as a function of
lookback time and redshift from Chary \& Elbaz (2001). {\bf b)}
Fraction of present-day stars ($+$ remnants) formed as a function of
lookback time and redshift for the scenario of
Fig.~\ref{FIG:mstar}a. Stellar lifetimes (for $Z_{\odot}$): Bressan et
al. (1993). Remnant masses: Prantzos \& Silk (1998).}
\label{FIG:mstar}
\end{figure}

\section{On the shyness of galaxies}  

The MIR-FIR correlations observed in the local universe (Elbaz et
al. 2002b) can be used to compute the contribution of the luminous IR
galaxies unveiled by ISOCAM below $z\sim$ 1.5 to the CIRB. Elbaz et
al. (2002b) computed a contribution of (16$\pm$5) nW m$^{-2}$
Hz$^{-1}$ as compared to the peak value of the CIRB of (25$\pm$7) nW
m$^{-2}$ Hz$^{-1}$ measured with COBE at $\lambda \sim$
140\,$\mu$m. Hence luminous IR galaxies below $z\sim$ 1.5 are
responsible for the bulk of the CIRB. Since the CIRB contains most
photons radiated by galaxies over the history of the universe, this
means that luminous IR galaxies represent a common phase for
galaxies. Chary \& Elbaz (2001) have studied the range of possible
parameters for the evolution of galaxies in luminosity and density
over the history of the universe, that would fit number counts from
ISOCAM, ISOPHOT and SCUBA as well as the CIRB and the redshift
distribution of ISOCAM galaxies. The major result of this study is
that although a level of degeneracy remains in the choice of the
parameters ruling the evolution of galaxies, existing observations set
a strong constraint on the relatively recent ($z<1.5$) evolution of
the number and luminosity density of luminous IR galaxies. The best
fit is obtained for the cosmic history of star formation shown in the
Fig.~\ref{FIG:mstar}a, where the relative roles of ULIGs, LIGs and
``normal'' galaxies are differentiated. Fig.~\ref{FIG:mstar}a implies
that we are living at an epoch when ``normal'' galaxies ($L_{\rm
bol}<$ 10$^{11}$ $L_{\odot}$) contribute dominantly to the global star
formation activity in the local universe, whereas above $z\sim$ 0.3
the reverse was true: the bulk of the cosmic density of star formation
was due to luminous IR galaxies. Hence galaxies in general must have
experienced a period of shyness, such as local LIGs and ULIGs, when
they formed the bulk of their present-day stars. Fig.~\ref{FIG:mstar}b
represents the fraction of present-day stars plus remnants formed as a
function of lookback time or redshift for a given IMF (here from Gould
et al. 1996). The total mass is comparable to the local density of
baryons in the local universe (5$\pm$3$\times 10^{8}~M_{\odot}$
Mpc$^{-3}$, Fukugita et al. 1998). The error bar on the computed
stellar mass is as large $\sim$ 50\,$\%$ (including the conversion
from MIR to FIR and FIR to SFR), but this result suggests that the
bulk of present-day stars formed at a time when their host galaxies
experienced such a phase of shyness 5 to 10 Gyr ago, i.e. between $z=$
0.5 and 2 for an age of the universe of 12.6 Gyr in our cosmology
(H$_o$= 75 km s$^{-1}$ Mpc$^{-1}$, $\Omega_{\rm matter}$= 0.3,
$\Omega_{\Lambda}= 0.7$). The shyness of galaxies seems to be the
result of galaxy encounters since all ISOCAM galaxies in the HDFN are
either merging or members of small groups of galaxies (see Aussel, in
this conference). The fact that the CIRB peaks around $\lambda\sim$
140\,$\mu$m was already an indication that it must originate from this
redshift range since galaxies SEDs peak above $\lambda \sim$
60\,$\mu$m.

\end{article}
\end{document}